\documentclass[review]{elsarticle}

\usepackage{hyperref}\usepackage{amsmath}

\journal{-}

\begin{document}

\begin{frontmatter}

\title{Cosmological isotropic matter-energy generalizations of Schwarzschild and Kerr metrics}

\author{Metin Arik}
\ead{metin.arik@boun.edu.tr}
\author{Yorgo Senikoglu}
\ead{yorgo.senikoglu@boun.edu.tr}
\address{Department of Physics, Bo\u{g}azi\c{c}i University, Bebek, Istanbul, Turkey}


\begin{abstract}
We present a time dependent isotropic fluid solution around a Schwarzschild black hole. We offer the solutions and discuss the effects on the field equations and the horizon. We derive the energy density, pressure and the equation of state parameter.
In the second part, we generalize the rotating black hole solution to an expanding universe. We derive from the proposed metric the special solutions of the field equations for the dust approximation and the dark energy solution. We show that the presence of a rotating black hole does not modify the scale factor $b(t)=t^{2/3}$ law for dust, nor $b(t)=e^{\lambda\hspace{1mm}t}$ and $p=-\rho$ for dark energy.  

\end{abstract}

\begin{keyword}
\texttt{Physics of Black Holes, General Formalism, Exact Solutions}
\PACS[2010]{04.20.Cv, 04.20.Jb, 04.70.-s}
\end{keyword}

\end{frontmatter}

\section{Introduction}
In order to describe relativistic spheres in general relativity, one must consider time dependent solutions. Earlier works for an expanding universe as exact solutions are widely known to us \cite{A},\cite{B}. Arakida\cite{C} presented the dominant effects due to cosmological expansion, and the use of a time dependent spacetime model.
Bhattacharya \textit{et al}., \cite{I} introduced a cosmological model that describes isotropic expansion of inhomogeneous universe,  and showed the energy-momentum tensor that creates the spatial inhomogeneity may not affect the uniform expansion scaling factor $a(t)$ in the FLRW-like metrics. Zhang \textit{et al}, \cite{J} obtained an exact time-dependent spherically symmetric solution describing gravitational collapse to a static scalar-hairy black hole. Joshi \textit{et al}., \cite{K} pointed that the black holes and naked singularities, which are consequences of the general theory of relativity as we consider the gravitational collapse of a massive star, would appear to be the leading candidates to explain the very high energy astrophysical phenomena being observed.
McVittie\cite{D},\cite{E}\hspace{1mm}incorporated two solutions to depict a spherically symmetric metric that describes a point mass embedded in an expanding spatially-flat universe.
The motivation that drove us to this part of the paper was the possibility if a solution satisfying only the isotropy condition existed in a non-static spacetime.

On the other hand, it is well known that the gravitational field of a rotating black hole is described by the Kerr metric\cite{F},\cite{P}. Vaidya \textit{et al}., \cite{G} have obtained an exact solution of
the Einstein's equations which describes the field of a radiating Kerr particle embedded in Einstein static universe. Vaidya \cite{H} has also given a very general form of the Kerr-Schild metric satisfying the Einstein's field equations.
Tzounis \textit{et al}., \cite{L} explored the properties of the ergoregion and the location of the curvature singularities for the Kerr black hole distorted by the gravitational field of external sources and they also studied the scalar curvature invariants of the horizon and compared their behaviour with the case of the isolated Kerr black hole. Lake \textit{et al}., \cite{M} examined the global structure of the family of Kerr-de Sitter spacetimes.  Abdelqader \textit{et al}., \cite{N} presented an invariant characterization of the physical properties of the Kerr spacetime  and introduced two dimensionless invariants, constructed out of some known curvature invariants, that act as detectors for the event horizon and ergosurface of the Kerr black hole. Gibbons  \textit{et al}., \cite{O} discussed the global structure of the metrics, and obtain formulae for the surface gravities and areas of the event horizons.
There is a considerable interest in generalizing the Kerr metric to describe the non-static field of a rotating star, that is why the main aim of the second part is to derive an exact solution of Einstein's equations which describes the field of a Kerr  source embedded in a rotating expanding universe.

The organization of this article is as follows. First we will derive a time dependent isotropic metric with an energy momentum tensor whose diagonal part is isotropic and contains a radial momentum and in the second part a time dependent parameter solution that generalizes the Kerr solution in an expanding universe.

\section{A cosmological generalization of the {S}chwarzschild black hole}

\noindent
We assume a form of the metric motivated by the Schwarzschild black hole in isotropic coordinates by replacing the mass parameter by an arbitrary function of r and t. Then we will impose the equality of the spatial diagonal elements of the energy-momentum tensor. Let us describe this by considering the following metric in a non-static spacetime generalized by a function $f(r,t)$ and a cosmological scale factor $a(t)$.
\begin{multline}
ds^2=\frac{(1-f(r,t))^2}{(1+f(r,t))^2}dt^2-a(t)^2(1+f(r,t))^4 (dr^2+r^2 d{\Omega}^2)\\where \hspace{2mm}d{\Omega}^2 = d{\theta}^2+\sin^2\theta d{\phi}^2.
\end{multline}
\noindent
The calculations lead us to the following non-vanishing components of the Einstein tensor in the orthonormal basis
${G}_{tt}$, ${G}_{tr}$, ${G}_{rr}$ and ${G}_{\theta\theta}={G}_{\phi\phi}$.

\noindent
We will not bother the reader with the long and exhausting details of these components at this point.
\noindent
The only condition that we impose is the following, in the orthonormal basis:
\begin{equation}
{G}_{rr}={G}_{\theta\theta}={G}_{\phi\phi}.
\end{equation}
The non-vanishing components of the Einstein tensor are presented in the {A}ppendix {A}.
\noindent
The unique solution that satisfies (2) is:
\begin{equation}
f(r,t)=\frac{1}{\sqrt{m(t)^2r^2+2n(t)}}. 
\end{equation}

\noindent
We notice that at $r=r_{H}$ we have $f(r,t)=1$, where $r_{H}$ is the horizon. 
\newpage
We denote p and $\rho$ respectively the pressure and density.
As $r\rightarrow\ r_{H}$, we set $\dot{f}(r,t)\rightarrow\ 0$ so that $\frac{p}{\rho}$ is non-singular at the horizon, this gives:
\begin{equation}
f(r,t)=\frac{1}{\sqrt{m(t)^2(r^2-r^2_{H})+1}}. 
\end{equation} 
\noindent
With (1) and (4) the non-vanishing components of the Einstein tensor in the orthonormal basis are presented in the {A}ppendix {B}.

\noindent
We note that as $r\rightarrow\ \infty$ we have:
\begin{equation}
{G}_{tt} = 3\frac{\dot{a}^2}{a^2}\hspace{2mm} and \hspace{2mm} {G}_{rr}={G}_{\theta\theta}={G}_{\phi\phi} = -2\frac{\ddot{a}}{a}-\frac{\dot{a}^2}{a^2}.
\end{equation}
And as $r\rightarrow\ r_{H}$:
\begin{equation}
{G}_{tt} = \frac{12}{a^2(1-f)^2}(\dot{f}a+\dot{a})^2 \hspace{2mm} and \hspace{2mm} {G}_{rr}={G}_{\theta\theta}={G}_{\phi\phi} = -\frac{8}{a^2(1-f)^3}a\dot{f}(\dot{f}a+\dot{a}).
\end{equation}
For further purposes we denote the equation of state parameter:
\begin{equation*}
\nu=\frac{p}{\rho}=-\frac{2}{3(1-f)}\frac{a\dot{f}}{\dot{f}a+\dot{a}}
\end{equation*}
with ${G}_{rr}={G}_{\theta\theta}={G}_{\phi\phi}=p$ and ${G}_{tt}=\rho$.

\noindent
What we observe for the equation of state parameter and the function m(t) as $r\rightarrow\ r_{H}$ is:
\begin{equation}
m^2(t)=-\frac{3\nu}{2r^2_{H}}ln(a(t)).
\end{equation}
We can see that for this solution to exist for an expanding universe we need $\nu\leq 0$. e.g. the dark energy solution $p=-\rho$.
\newpage
\section{A cosmological generalization of the {K}err black hole}
The Kerr metric describes the geometry of spacetime in the vicinity of a mass $M$ rotating with angular momentum $J$. We add to this metric in Boyer-Lindquist coordinates a scale factor that we denote $b(t)$.
The metric can be written as follows:
\begin{multline}
ds^2=-(1-\frac{2Mr}{\rho^2})dt^2-b(t)\big(\frac{4Mr\alpha sin^2\theta}{\rho^2}\big)d{\phi}dt+b(t)^2\big(\frac{\rho^2}{\Delta}\big)dr^2 \\ +b(t)^2(\rho^2)d{\theta}^2+b(t)^2(r^2+\alpha^2+\frac{2Mr\alpha^2sin^2\theta}{\rho^2})\sin^2\theta d{\phi}^2.
\end{multline}
where $\alpha=\frac{J}{M}$, $\rho^2=r^2+\alpha^2cos^2\theta$ and $\Delta=r^2-2Mr+\alpha^2$.
\noindent
The non-vanishing components of the Einstein Tensor in the orthonormal basis are:
${G}_{tt}$, ${G}_{rr}={G}_{\theta\theta}={G}_{\phi\phi}$, ${G}_{tr}$,${G}_{t\theta}$, ${G}_{t\phi}$, ${G}_{r\phi}$ and ${G}_{\theta\phi}$.
\paragraph{Dust approximation}
Matter dominated Universe is modeled by dust approximation. The matter is approximated as dust particles which produce no pressure.
Solving for:
\begin{equation}
{G}_{rr}={G}_{\theta\theta}={G}_{\phi\phi}=0.
\end{equation}
we obtain:
\begin{equation}
b(t)=b_{1}t^{2/3}
\end{equation}
where $b_{1}$ is a constant.
Now for the non-vanishing components of the Einstein Tensor in the orthonormal basis we get:
\begin{multline}
{G}_{tt}=\frac{4}{3t^2}\hspace{1mm}\frac{\rho^2}{r^2-2Mr+\alpha^2cos^2\theta}, {G}_{tr}=-\frac{4}{3b_{1}t^{5/3}}\frac{M(\alpha^2+r^2)(\alpha^2cos^2\theta-r^2)}{\sqrt{\Delta(r^2-2Mr+\alpha^2cos^2\theta)}\rho^4} \\ 
{G}_{t\theta}=-\frac{4}{3b_{1}t^{5/3}}\frac{Mr\alpha^2sin2\theta}{\sqrt{r^2-2Mr+\alpha^2cos^2\theta)}\rho^4}, {G}_{t\phi}=0 \\ 
{G}_{r\phi}=-\frac{4}{3b_{1}t^{5/3}}\frac{\alpha Msin^2\theta(3r^4+\alpha^2r^2+\alpha^2r^2cos^2\theta-\alpha^4cos^2\theta)}{\rho^4\sqrt{\Delta sin^2\theta((\alpha^2cos^2\theta)\Delta+r^4+2Mr\alpha^2+\alpha^2r^2)}}\hspace{2mm} and \\ {G}_{\theta\phi}=\frac{4}{3b_{1}t^{5/3}}\frac{\alpha^3Mr(sin2\theta) (sin^2\theta)}{\rho^4 \sqrt{sin^2\theta ((\alpha^2cos^2\theta)\Delta+r^4+2Mr\alpha^2+\alpha^2r^2)}}.
\end{multline}
\paragraph{Dark Energy}
For a dark energy dominated universe let us solve the equation:
\begin{equation}
{G}_{tt}=-{G}_{rr}=-{G}_{\theta\theta}=-{G}_{\phi\phi}.
\end{equation}
we obtain the solution:
\begin{equation}
b(t)=b_{0}e^{\sqrt{\frac{\Lambda}{3}}\hspace{1mm}t}.
\end{equation}
where $b_{0}$ is a constant.
The components of the Einstein Tensor in the orthonormal basis are:
\begin{multline}
{G}_{tt}=-{G}_{rr}=-{G}_{\theta\theta}=-{G}_{\phi\phi}=\frac{\Lambda\big((\Delta)(\alpha^2cos^2\theta)+r^4+2Mr\alpha^2+\alpha^2r^2\big)}{\rho^2 \hspace{2mm} \Delta}, \\ {G}_{tr}=-\frac{2}{3b_{0}}\frac{M(\alpha^2+r^2)(\alpha^2cos^2\theta-r^2)(\sqrt{3\Lambda}e^{-\sqrt{\frac{\Lambda}{3}}\hspace{1mm}t})}{\sqrt{\Delta(r^2-2Mr+\alpha^2cos^2\theta)}\rho^4}, \\ {G}_{t\theta}=-\frac{2}{3b_{0}}\frac{(Mr\alpha^2sin2\theta)(\sqrt{3\Lambda}e^{-\sqrt{\frac{\Lambda}{3}}\hspace{1mm}t})}{\sqrt{r^2-2Mr+\alpha^2cos^2\theta}\rho^4}, \\ {G}_{t\phi}=\frac{(2\alpha Mr\Lambda sin^2\theta)((\alpha^2cos^2\theta)\Delta + r^4+2Mr\alpha^2+\alpha^2r^2)}{\rho^2\Delta \sqrt{sin^2\theta ((\alpha^2cos^2\theta)\Delta+r^4+2Mr\alpha^2+\alpha^2r^2)(r^2-2Mr+\alpha^2cos^2\theta)}}, \\ {G}_{r\phi}=-\frac{2}{3b_{0}}\frac{\alpha Msin^2\theta (3r^4+\alpha^2r^2+\alpha^2r^2cos^2\theta-\alpha^4cos^2\theta)(\sqrt{3\Lambda}e^{-\sqrt{\frac{\Lambda}{3}}\hspace{1mm}t})}{\rho^4\sqrt{\Delta sin^2\theta((\alpha^2cos^2\theta)\Delta+r^4+2Mr\alpha^2+\alpha^2r^2)}}\hspace{2mm} and \\ {G}_{\theta\phi}=\frac{2}{3b_{0}}\frac{\alpha^3Mr(sin2\theta) (sin^2\theta)(\sqrt{3\Lambda}e^{-\sqrt{\frac{\Lambda}{3}}\hspace{1mm}t})}{\rho^4 \sqrt{sin^2\theta ((\alpha^2cos^2\theta)\Delta+r^4+2Mr\alpha^2+\alpha^2r^2)}}.
\end{multline}
\section{Discussion of Solutions}
For the Schwarzschild black hole solution that we have presented, a diagonal energy-momentum tensor satisfying ${G}_{rr}={G}_{\theta\theta}={G}_{\phi\phi}$ corresponds to an isotropic and homogeneous fluid. An energy momentum tensor satisfying ${G}_{rr}={G}_{\theta\theta}={G}_{\phi\phi}$ with the only non-diagonal non-vanishing term ${G}_{tr}$ means that we can still talk of a pressure but there is a non-vanishing radial momentum density corresponding to the infalling matter.

For the Kerr black hole, it is desirable to introduce a mass $M$ rotating with angular momentum $J$ where both $J$ and $M$ are time dependent but in this case the equations do not yield any solution. By calculating the energy-momentum tensor, we have obtained some off-diagonal terms ${G}_{tr}, {G}_{t\theta}, {G}_{t\phi}$ which are momentum density components, and some shear stress like ${G}_{r\phi}$ and ${G}_{\theta\phi}$. We have presented our work as a first attempt to describe also these shortcomings.

On the other hand, it is well-known that introducing a cosmological constant is equivalent to adding a part to the energy-momentum tensor satisfying ${G}_{tt}=-{G}_{rr}=-{G}_{\theta\theta}=-{G}_{\phi\phi}$. The solutions that we get by solving the Einstein equations with a cosmological constant are the same; only the interpretation of the energy-momentum tensor is different. In our presentation we choose a language where there is no cosmological constant but where a term satisfying ${G}_{tt}=-{G}_{rr}=-{G}_{\theta\theta}=-{G}_{\phi\phi}$ in the energy-momentum tensor is allowed.

In the recent observations that have demonstrated the existence of a binary stellar-mass black hole system\cite{Q} and the first observation of a binary black hole merger, it has been pointed out that the angular momentum observed was very high. This may support a detail of our work concerning the angular momentum components that we have noted in calculating the Einstein tensor.
Since we believe that there is a black hole at the center of each galaxy and that the rotation axes of galaxies are not isotropically distributed in space, an isotropic and homogeneous cosmology does not seem to be realistic.
 
\section{Conclusion}
We have studied the effects of a time dependent isotropic metric with an energy momentum tensor whose diagonal part is isotropic and contains a radial momentum. Consequently, field equations were presented for various cases to illustrate our point of view. We have derived the energy, pressure and the equation of state parameter $\nu$ that satisfied the condition ${G}_{rr}={G}_{\theta\theta}={G}_{\phi\phi}$ and obtained a non-singular expression for $\frac{p}{\rho}$ that binds the scale factor, $r_{H}$, $\nu$ and the function figuring in the metric proposed for the latter condition.   

From a more generalized {K}err metric in {B}oyer-{L}indquist coordinates that we have proposed in this paper, we have shown that, depending on given conditions, we can obtain the characteristics of the scale factor $b(t)$ for the matter dominated universe ($p=0$, matter dominated dust approximation) and the vacuum (dark) energy dominated universe ($p=-\rho$). We notice the known results that for $\alpha\rightarrow 0$ the {B}oyer-{L}indquist line element reproduces the {S}chwarzschild line element in an expanding universe and for $M\rightarrow 0$ the flat expanding {M}inkowski space.

For a matter dominated universe, we observe ${G}_{t\phi}=0$ indicating no angular momemtum in the azimuthal direction. We note that as $r\rightarrow \infty$, ${G}_{tt}$ is independent of r and ${G}_{tt}\sim\frac{1}{t^2}$ , ${G}_{tr}\sim\frac{1}{r^2t^{\frac{5}{3}}}$, ${G}_{t\theta}\sim\frac{1}{r^4t^{\frac{5}{3}}}$, ${G}_{r\phi}\sim\frac{1}{r^3t^{\frac{5}{3}}}$ and ${G}_{\theta\phi}\sim\frac{1}{r^5t^{\frac{5}{3}}}$.
For a dark energy dominated universe, we see that ${G}_{tt}$, ${G}_{rr}$, ${G}_{\theta\theta}$, ${G}_{\phi\phi}$ and  ${G}_{t\phi}$ are independent of time. ${G}_{tr}\sim\frac{1}{r^4e^{\sqrt{\frac{\Lambda}{3}}\hspace{1mm}t}}$, ${G}_{t\theta}\sim\frac{1}{r^4e^{\sqrt{\frac{\Lambda}{3}}\hspace{1mm}t}}$, ${G}_{t\phi}\sim\frac{1}{r^2}$, ${G}_{r\phi}\sim\frac{1}{r^3e^{\sqrt{\frac{\Lambda}{3}}\hspace{1mm}t}}$ and ${G}_{\theta\phi}\sim\frac{1}{r^5e^{\sqrt{\frac{\Lambda}{3}}\hspace{1mm}t}}$.

\newpage
\section{Appendices}
\paragraph{Appendix A}
\begin{multline*}
ds^2=\frac{(1-f(r,t))^2}{(1+f(r,t))^2}dt^2-a(t)^2(1+f(r,t))^4 (dr^2+r^2 d{\Omega}^2)
\hspace{2mm}where \hspace{2mm}d{\Omega}^2 = d{\theta}^2+\sin^2\theta d{\phi}^2.
\end{multline*}
We obtain the following non-vanishing components of the Einstein tensor in the orthonormal basis:
\begin{multline*}
{G}_{tt}=\frac{1}{(1+f)^5(1-f)^2}\bigg(3\frac{\dot{a}^2}{a^2}(f^7)+(12\frac{\dot{a}}{a}\dot{f}+21\frac{\dot{a}^2}{a^2})(f^6)+(12\dot{f}^2+72\frac{\dot{a}}{a}\dot{f}+63\frac{\dot{a}^2}{a^2})(f^5)\\ +(60\dot{f}^2+180\frac{\dot{a}}{a}\dot{f}+105\frac{\dot{a}^2}{a^2})(f^4)+(120\dot{f}^2+240\frac{\dot{a}}{a}\dot{f}+105\frac{\dot{a}^2}{a^2})(f^3)\\ +(120\dot{f}^2+180\frac{\dot{a}}{a}\dot{f}+63\frac{\dot{a}^2}{a^2}-4\frac{{f}_{r,r}}{a^2}-8\frac{{f}_{r}}{ra^2})(f^2)+(60\dot{f}^2+72\frac{\dot{a}}{a}\dot{f}+21\frac{\dot{a}^2}{a^2}+8\frac{{f}_{r,r}}{a^2}+16\frac{{f}_{r}}{ra^2})(f)\\ +(12\dot{f}^2+12\frac{\dot{a}}{a}\dot{f}+3\frac{\dot{a}^2}{a^2}-4\frac{{f}_{r,r}}{a^2}-8\frac{{f}_{r}}{ra^2})\bigg)\\ 
{G}_{tr}=-\frac{4a{f}_{r}{f}_{t}+a{f}_{r,t}(1-f)+\dot{a}{f}_{r}}{a^2(1+f)^2(1-f)^2}\\
{G}_{rr}=-\frac{1}{(1+f)^5(1-f)^3}\bigg((2\frac{\ddot{a}}{a}+\frac{\dot{a}^2}{a^2})(f^8)+(4\ddot{f}+12\frac{\dot{a}}{a}\dot{f}+12\frac{\ddot{a}}{a}+6\frac{\dot{a}^2}{a^2})(f^7)+(8\dot{f}^2+20\ddot{f}+56\frac{\dot{a}}{a}\dot{f}+28\frac{\ddot{a}}{a}+14\frac{\dot{a}^2}{a^2})(f^6)\\ +(24\dot{f}^2+36\ddot{f}+84\frac{\dot{a}}{a}\dot{f}+28\frac{\ddot{a}}{a}+14\frac{\dot{a}^2}{a^2})(f^5)+(20\ddot{f})(f^4)+(-80\dot{f}^2-20\ddot{f}-140\frac{\dot{a}}{a}\dot{f}-28\frac{\ddot{a}}{a}-14\frac{\dot{a}^2}{a^2}-4\frac{{f}_{r}}{ra^2})(f^3)\\ +(-120\dot{f}^2-36\ddot{f}-168\frac{\dot{a}}{a}\dot{f}-28\frac{\ddot{a}}{a}-14\frac{\dot{a}^2}{a^2}+8\frac{{f}_{r}}{ra^2}-4\frac{{f}_{r}^2}{a^2})(f^2)+(-72\dot{f}^2-20\ddot{f}-84\frac{\dot{a}}{a}\dot{f}-12\frac{\ddot{a}}{a}-6\frac{\dot{a}^2}{a^2}-4\frac{{f}_{r}}{ra^2}+8\frac{{f}_{r}^2}{a^2})(f)\\ +(-16\dot{f}^2-4\ddot{f}-16\frac{\dot{a}}{a}\dot{f}-2\frac{\ddot{a}}{a}-\frac{\dot{a}^2}{a^2}-4\frac{{f}_{r}^2}{a^2})\bigg)\\
{G}_{\theta\theta}={G}_{\phi\phi}=-\frac{1}{(1+f)^5(1-f)^3}\bigg((2\frac{\ddot{a}}{a}+\frac{\dot{a}^2}{a^2})(f^8)+(4\ddot{f}+12\frac{\dot{a}}{a}\dot{f}+12\frac{\ddot{a}}{a}+6\frac{\dot{a}^2}{a^2})(f^7)\\ +(8\dot{f}^2+20\ddot{f}+56\frac{\dot{a}}{a}\dot{f}+28\frac{\ddot{a}}{a}+14\frac{\dot{a}^2}{a^2})(f^6)\\ +(24\dot{f}^2+36\ddot{f}+84\frac{\dot{a}}{a}\dot{f}+28\frac{\ddot{a}}{a}+14\frac{\dot{a}^2}{a^2})(f^5)+(20\ddot{f})(f^4)+(-80\dot{f}^2-20\ddot{f}-140\frac{\dot{a}}{a}\dot{f}-28\frac{\ddot{a}}{a}-14\frac{\dot{a}^2}{a^2}-2\frac{{f}_{r}}{ra^2}-2\frac{{f}_{rr}}{a^2})(f^3)\\ +(-120\dot{f}^2-36\ddot{f}-168\frac{\dot{a}}{a}\dot{f}-28\frac{\ddot{a}}{a}-14\frac{\dot{a}^2}{a^2}+4\frac{{f}_{r}}{ra^2}+2\frac{{f}_{r}^2}{a^2}+4\frac{{f}_{rr}}{a^2})(f^2)\\ +(-72\dot{f}^2-20\ddot{f}-84\frac{\dot{a}}{a}\dot{f}-12\frac{\ddot{a}}{a}-6\frac{\dot{a}^2}{a^2}-2\frac{{f}_{r}}{ra^2}-4\frac{{f}_{r}^2}{a^2}-2\frac{{f}_{rr}}{a^2})(f)\\ +(-16\dot{f}^2-4\ddot{f}-16\frac{\dot{a}}{a}\dot{f}-2\frac{\ddot{a}}{a}-\frac{\dot{a}^2}{a^2}+2\frac{{f}_{r}^2}{a^2})\bigg)\\
\end{multline*}
\newpage
\paragraph{Appendix B}
$f(r,t)=\frac{1}{\sqrt{m(t)^2(r^2-r^2_{H})+1}}$
\begin{multline*}
{G}_{tr}=\frac{4}{a^2(1+f)^2(1-f)^2}\bigg(f^3(2mr\dot{m}a+\dot{a}^2m^2r)+f^4(-2m\dot{m}ra)+f^5(-3m^3r(r^2-r^2_{H})\dot{m}a)+ \\ f^6(2m^3r(r^2-r^2_{H})\dot{m}a)\bigg) \\ 
{G}_{tt}=\frac{1}{(1+f)^5(1-f)^2}\bigg(3\frac{\dot{a}^2}{a^2}+(21\frac{\dot{a}^2}{a^2})f+(63\frac{\dot{a}^2}{a^2})f^2+(105\frac{\dot{a}^2}{a^2}-12\frac{\dot{a}}{a}m\dot{m}(r^2-r^2_{H})+12\frac{m^2}{a^2})f^3 \\ +(105\frac{\dot{a}^2}{a^2}-72\frac{\dot{a}}{a}m\dot{m}(r^2-r^2_{H})-24\frac{m^2}{a^2})f^4+(63\frac{\dot{a}^2}{a^2}-180\frac{\dot{a}}{a}m\dot{m}(r^2-r^2_{H})+12\frac{m^2}{a^2}-12\frac{r^2m^4}{a^2})f^5 \\ +(21\frac{\dot{a}^2}{a^2}-240\frac{\dot{a}}{a}m\dot{m}(r^2-r^2_{H})+12m^2\dot{m}^2(r^4+r^4_{H})-24r^2m^2\dot{m}^2r^2_{H}+24\frac{r^2m^4}{a^2})f^6 \\ +(3\frac{\dot{a}^2}{a^2}-180\frac{\dot{a}}{a}m\dot{m}(r^2-r^2_{H})+60m^2\dot{m}^2(r^4+r^4_{H})-120r^2m^2\dot{m}^2r^2_{H}-12\frac{r^2m^4}{a^2})f^7 \\ +(-72\frac{\dot{a}}{a}m\dot{m}(r^2-r^2_{H})-240r^2m^2\dot{m}^2r^2_{H}+120m^2\dot{m}^2(r^4+r^4_{H}))f^8 \\ +(-12\frac{\dot{a}}{a}m\dot{m}(r^2-r^2_{H})-240r^2m^2\dot{m}^2r^2_{H}+120m^2\dot{m}^2(r^4+r^4_{H}))f^9 \\ +(-120r^2m^2\dot{m}^2r^2_{H}+60m^2\dot{m}^2(r^4+r^4_{H}))f^{10} \\ +(-24r^2m^2\dot{m}^2r^2_{H}+12m^2\dot{m}^2(r^4+r^4_{H}))f^{11}\bigg)\hspace{2mm} and \\ 
{G}_{rr}={G}_{\theta\theta}={G}_{\phi\phi}=-\frac{1}{(1+f)^5(1-f)^3}\bigg(2\frac{\ddot{a}}{a}+\frac{\dot{a}^2}{a^2}+6(2\frac{\ddot{a}}{a}+\frac{\dot{a}^2}{a^2})f+14(2\frac{\ddot{a}}{a}+\frac{\dot{a}^2}{a^2})f^2 \\ +(14(2\frac{\ddot{a}}{a}+\frac{\dot{a}^2}{a^2})-4(r^2-r^2_{H})(\ddot{m}m+\dot{m}^2+4m\dot{m}\frac{\dot{a}}{a}))f^3 \\ -4((r^2-r^2_{H})(5\ddot{m}m+5\dot{m}^2+21m\dot{m}\frac{\dot{a}}{a})+\frac{m^2}{a^2})f^4+ \\ (-14(2\frac{\ddot{a}}{a}+\frac{\dot{a}^2}{a^2})-4(r^2-r^2_{H})(9\ddot{m}m+9\dot{m}^2+42m\dot{m}\frac{\dot{a}}{a})+12m^2\dot{m}^2(r^4+r^4_{H})-24m^2\dot{m}^2r^2r^2_{H}+8\frac{m^2}{a^2})f^5 \\ +(-14(2\frac{\ddot{a}}{a}+\frac{\dot{a}^2}{a^2})-20(r^2-r^2_{H})(\ddot{m}m+\dot{m}^2+7m\dot{m}\frac{\dot{a}}{a})+76m^2\dot{m}^2(r^4+r^4_{H})-152m^2\dot{m}^2r^2r^2_{H}+4\frac{m^2}{a^2}(m^2r^2+1))f^6 \\ +(-6(2\frac{\ddot{a}}{a}+\frac{\dot{a}^2}{a^2})-20(r^2-r^2_{H})(\ddot{m}m+\dot{m}^2)+180m^2\dot{m}^2(r^4+r^4_{H})-360m^2\dot{m}^2r^2r^2_{H}-8\frac{r^2m^4}{a^2})f^7 \\ +(-(2\frac{\ddot{a}}{a}+\frac{\dot{a}^2}{a^2})+4(r^2-r^2_{H})(9\ddot{m}m+9\dot{m}^2+21m\dot{m}\frac{\dot{a}}{a})+180m^2\dot{m}^2(r^4+r^4_{H})-360m^2\dot{m}^2r^2r^2_{H}+4\frac{r^2m^4}{a^2})f^8 \\ +(4(r^2-r^2_{H})(5\ddot{m}m+5\dot{m}^2+14m\dot{m}\frac{\dot{a}}{a})+20m^2\dot{m}^2(r^4+r^4_{H})-40m^2\dot{m}^2r^2r^2_{H})f^9 \\ +(4(r^2-r^2_{H})(\ddot{m}m+\dot{m}^2+3m\dot{m}\frac{\dot{a}}{a})-108m^2\dot{m}^2(r^4+r^4_{H})-216m^2\dot{m}^2r^2r^2_{H})f^{10} \\ +(168m^2\dot{m}^2r^2r^2_{H}-84m^2\dot{m}^2(r^4+r^4_{H}))f^{11}+(40m^2\dot{m}^2r^2r^2_{H}-20m^2\dot{m}^2(r^4+r^4_{H}))f^{12}\bigg).
\end{multline*}

\bibliography{m4}

\end{document}